\documentstyle[11pt]{article}

 \csname
@addtoreset\endcsname{equation}{section}

\def\a{\alpha}

\def\d{\delta}

\def\e{\epsilon}

\def\l{\lambda}

\def\m{\mu}
\def\n{\nu}

\def\s{\sigma}

\def\th{\theta}

\def\x{\xi}

    \def\ra{\rightarrow}

\def\del{\partial}
\let\la=\label
\let\bm=\bibitem

\newcommand{\eq}[1]{(\ref{#1})}

\def\be{\begin{equation}}
\def\ee{\end{equation}}
\def\bea{\begin{eqnarray}}
\def\eea{\end{eqnarray}}
\def\ba{\begin{array}}
\def\ea{\end{array}}

\def\ed{\end{document}}

\newcommand{\hoch}[1]{$\, $}

\newcommand{\tamphys}{
\begin{center}
{\it Center for Theoretical Physics,
Texas A\&M University, College Station, TX 77843, USA}
\end{center}
}

\newcommand{\auth}{\large  I. Rudychev
\hoch{1}}

\begin{document}

\hfill{CTP-TAMU-02/01}

\hfill{hep-th/0101039}

\hfill{\today}

\vspace{20pt}

\begin{center}


{\Large \bf From noncommutative string/membrane to ordinary ones.}


\vspace{30pt}


\auth

\vspace{15pt}

\begin{itemize}

\item[$ $] \tamphys

\end{itemize}

\vspace{30pt}

{\bf Abstract}

\end{center}

We discuss origin of equivalence between noncommutative and ordinary 
Yang-Mills from point of view of string theory. Working in BRST/Hamiltonian
framework first we investigate string model in the decoupling limit and show
that change of variables and applying the conversion of constraints
of decoupled string theory gives commuting coordinates on the D-brane.
Also, we discuss algebra of constraints in general case and show the ways
of having commutative coordinates without going to decoupling limit. It could be
argued that noncommutative string in B-field is equivalent to the commutative
model. We investigate the case of the membrane ending on the M-5-brane
in constant C-field and discuss noncommutative/commutative equivalence
in this case.

{\vfill\leftline{}\vfill \vskip 10pt \footnoterule {\footnotesize
\hoch{\dagger} e-mail: rudychev@rainbow.physics.tamu.edu

\vskip -12pt}

\pagebreak

\setcounter{page}{1}


\section{Introduction.}

String theory appeared extremely useful in study of the properties of
non-commutative field theories. Moreover, considering strings and membranes
in constant B-field background not only helps to investigate properties
of the low-energy field theories \cite{Wit1} but also gives new insight on the string and 
M-theory itself by producing new models and connecting old ones for example
 into the framework of OM-theory \cite{OM1},\cite{SST},\cite{Berg1},\cite{Berg2}.
  From this point of view it is interesting to ask if there is correspondence 
between non-commutative and ordinary
Yang-Mills appearing as decopling limit of the string theory in the large
B-field, how the correspondence between both non-commutative and ordinary cases
will be looking like from the point of you of strings and membranes.\\
  In this note we study first the decoupling limits of both theories and 
investigate the possible ways of disappearing non-commutativity.
First we start from example of decoupling limit of bosonic string. Here we work
in the Hamiltonian formalism which is equivalent to world-sheet treatment and 
then discuss the procedure from both points of view.
Correspondence between non-commutative and ordinary Yang-Mills comes from 
considering instantons in both theories \cite{Wit1}. But how can we see that 
from point of view of the string theory.
From the world-sheet consideration non-commutativity appears from interpreting 
time ordering as operator ordering 
in two-point functions as well as in product of vertex operators \cite{Sch1}. 
In the Hamiltonian treatment non-commutativity arises from modification of the
usual constraints describing open string theory by additional constraint
coming from modified boundary conditions due to presence of the constant 
B-field \cite{Chu1}, \cite{Chu2}, \cite{Jab1} \cite{Jab2}, \cite{Oh1} and  \cite{Lee}.
 This new constraints, coming from boundary, is of the second class and 
for a first sight spoils commutational relations between the target space 
coordinate and momenta on the world-volume of D-brane. Apparently, one of the 
ways to struggle with the second class constraints is to introduce the Dirac 
brackets \cite{Chu2}, \cite{Jab2}, \cite{Oh1} and \cite{Lee}. On these Dirac 
brackets inconsistency between commutators of 
coordinates and momenta disappear, instead, target space coordinates of the 
string on the surface of D-brane do not commute anymore.\\ 
 There is another way of working with the second class constraints - to use the
conversion procedure \cite{Fad}, \cite{Manv}, \cite{BF1}.
 Usually, conversion means extending the phase space of variables
by introducing new degrees of freedom as well as modification of the second 
class constraints in such a way that they become of the first class, i.e. 
either commute or give another constraints. This new first class constraints correspond, 
as usually, to the gauge symmetries of the theory, i.e. by introducing new
variables we acquire new gauge symmetries of the generalized action.
This framework could be understood from BRST-quantization point of view.
In this case one can treat additional variables as  ghosts and new modified action in 
additional to previous symmetries possess new gauge invariance. 
It is similar to having new local fermionic symmetry in ordinary BRST approach.
But in the case of the conversion new symmetry could be bosonic.
There were interesting connections of BRST conversion procedure with Fedosov quantization
\cite{Lyakh}.\\
 We apply conversion to the system with second class constraints and instead of going to 
Dirac brackets we end up with the model with all or crucial part
of constraints of the first class. In this case we don't have Dirac brackets
and all target space coordinates commute between each other.
In some sense one is able to substitute model with non-commutative 
coordinates by the ordinary ones on the level of BRST quantization. It is not necessary 
to BRST quantize the theory, it is possible to use ordinary Dirac 
quantization of systems with first class constraints only, but we need the former one 
to get rid of ambiguity of appearance of the new variables and to interpret these extra 
coordinates of the extended phase space as a ``ghosts''.
Here we mostly consider bosonic part of the theories.
It could be extended to the supersymmetric
case by following \cite{CZ}, \cite{LW}, \cite{LZ} and references therein.
Description of connection of noncommutative Yang-Mills with ordinary one
was also considered in \cite{JS}.
 
The organization of paper is the following:\\
 In section 2 we start from decoupling limit of the string in constant 
B-field and then see origin of non-commutativity from Hamiltonian analysis of
the constraints using Dirac brackets. Then we apply conversion to the system, 
finding equivalent one with commutative coordinates.\\
 In section 3 we discuss general framework of the bosonic string in the
B-field, consider origin of non-commutativity and discuss possible ways of its
removal.\\
 In section 4 we investigate decoupling limit of the membrane ending on
M-5-brane in large C-field. We discuss analysis of constraints and ways of 
obtaining equivalent theory without non-commutativity.\\
Discussions and conclusion could be found in section 5.\\


\section{String in constant B-field, decoupling limit and constraints. }
 
 The bosonic part of the world-sheet  action of the string ending on D-brane
 in ten dimensional Minkowski space in constant B-field background is given by

\be
S = \frac {1}{4\pi \a'} \int_{ M^2}^{ab}\del X^\m \del X_\m 
   - \frac{1}{2 \pi \a'} \int_{\del M^2} B_{ab}X^a \del_t X^b,
\la{S1}
\ee

where  $B_{ab}$ is nonzero only along Dp-brane and in general case includes B-field
together with two-form field strength on the D-brane. The boundary conditions
along D-brane are

\be
 g_{ab} \del_n X^b + 2 \pi \a' B_{ab} \del_t X^b = 0,
\la{bound1}
\ee

where $\del_n$ is derivative normal to the boundary of the world-sheet.

Here the effective metric seen by the open string is

\be
G^{ab} = \Biggl ( \frac {1}{g+2\pi \a'B}  g  \frac {1}{g-2 \pi \a' B} \Biggr )
^{ab},
\la{G1}
\ee

\be
G_{ab} = g_{ab} - (2 \pi \a')^2 (B  g^{-1} B)_{ab}.
\la{G2}
\ee

And noncommutativity parameter

\be
 \th^{ab} =- (2 \pi \a')^2 \Biggl ( \frac {1}{g+2 \pi \a'} B \frac {1}{g-2 \pi 
\a' B} \Biggr )^{ab}.
\la{th}
\ee

In the decoupling, zero slope, limit \cite{Wit1} one can take $\a' \ra 0$ keeping fixed
open string parameters, i.e. on the world volume of D-brane

\be
G^{ab} = - \frac{1}{(2 \pi \a')^2} \biggl ( \frac {1}{B} g \frac {1}{B}
 \biggr )^{ab},
\nonumber
\la{G3}
\ee

\be
G_{ab} = - (2 \pi \a')^2 (B g^{-1} B)_{ab},
\la{G4}
\ee

\be
\th^{ab} =  \biggl( \frac {1}{B} \biggr )^{ab}.
\nonumber
\la{th2}
\ee

In this limit the kinetic term in \eq{S1} vanishes. The remaining part, which
governs the dynamics is the second term in the \eq{S1} which in this case
describes the evolution of the boundary on the string, i.e. particle living
on the world-volume of D-brane.
This part of the action is given by

\be
S = \frac {1}{2} \int_{\del M^2} dt B_{ab} X^a \del_t X^b.
\la{Slim}
\ee

For simplicity and without loss of generality one can take $B_{ab}$ nonzero only in two space directions on 
the D-brane, i.e. $B_{ab} = 2b \e_{ab}$ where $a,b = 1,2$ and the rest of string coordinates
do not give any contribution to \eq{Slim}.

Now let us see how noncommutativity appears on the level of Hamiltonian 
formalism. The string action in decoupling limit is

\be
S = b \int dt \e_{ab} X^a \dot X^b.
\la{Sb}
\ee

Canonical momentum is given by

\be
 P_a = -b \e_{ab}X^b.
\la{PP}
\ee

So, the constraints that fully describe this model are

\be
\phi_a = P_a + b \e_{ab} X^b.
\la{constrP}
\ee

All these constraints are of the second class, i.e. they don't commute with
each other:

\be
[\phi_a, \phi_b] = 2b \e_{ab}.
\la{phiP}
\ee

Going to Dirac brackets gives

\be
[X^a,X^b]_D = [X^a,X^b]  - [X^a, \phi^c] \frac {\e^{cd}}{2b} [\phi^d, X^b],
\la{XX1}
\ee

and 

\be
[X^a, X^b]_D = \frac {1}{2b}\e^{ab}.
\la{XX2}
\ee

We see that coordinates of the string on the D-brane do not commute
and in the decoupling limit one has noncommutative Yang-Mills on the world-volume of D-branes.
It is possible to calculate a two-point function of the fields propagating on the boundary of the
string world-sheet. And it is given by

\be
< X^a (t), X^b (0) > = \frac {1}{2b} \e^{ab},
\la{prop}
\ee

which leads to noncommutative coordinates on the D-brane.

Another way to treat constraints \eq{constrP} is to extend the phase space of the
variables and in the framework of BRST quantization to introduce additional pair
of canonically conjugated variables $\x$ and $ P_{\x}$ with commutator
 $[\x, P_{\x}]  = 1$.
In this case it is possible to convert the second class constraints \eq{constrP} into the first one.

\be
 \hat \phi_1 = P_1 + b (X^2 - \sqrt{2}  \x) , \qquad \hat \phi_2 = (P_2 + \sqrt 2  P_{\x}) - b X^1,
\la{Chi-hat}
\ee

where new constraints are of the first class and they commute.
 Let us count number of degrees of freedom to be sure that we did not loose
any information. 
Each of the second class constraints eliminate one degree of freedom, but
each of the first class one kills two, so instead of two second class
constraints we have two first one now, therefore we need to add two independent degrees of freedom 
(i.e. $\x$ and $ P_{\x}$ ).

Now we can identify new variables, they are: $P_1, X_1$ which are unchanged and 

\be
P^{+} = -P_2 - \sqrt 2  P_{\x}, \qquad X^{ -} = X^2 - \sqrt 2 \x,
\la{P+}
\ee

and 

\be
P^{-} = -P_2 + \sqrt 2  P_{\x}, \qquad X^{ +} = X^2 + \sqrt 2 \x,
\la{P+1}
\ee

where they are consequently canonically conjugate. The variables $ P^{-}, X^{+}$
are not dynamical ones because they don't participate in the constraints.
\eq{Chi-hat}  now is taking the form:

\be
\hat\phi_1 = P_1 + b X^{-}, \qquad  \hat \phi_2 = P^{+} + b X^1.
\la{constr-P}
\ee

Those two constraint not only of the first class but also assume that propagator \eq{prop} is not
antisymmetric anymore, but rather symmetric in interchange of $ a$ and $b$ and leads to commutative 
coordinates not only on the level of Poisson brackets but on the level of two-point function on the
boundary of the string worldsheet, that could be seen using correspondence between operator and time
ordering \cite{Sch1}. So, new two-point functions are

\be
< X^1 (t) , X^{-} (0) >  = < X^{-}(t), X^1(0) >,
\la{prop1}
\ee

and could be calculated using path-integral BRST approach.
Therefore, we  removed the noncommutativity of the  string end-point coordinates by introducing
new
 variables, and applying the conversion of the constraints. By this way we  ended up with only first
class
constraints
where we don't have to use Dirac brackets and all coordinates commute on the level of the Poisson
brackets.
Also, after field redefinition the new coordinates of the string boundary are described now by  
$X^1  , X^{-}$, that commute, because propagator is symmetric. So using time ordering product gives
commutativity on the boundary of string world-sheet. 
It could be more appropriate showed by using BRST quantization of the theory described by the constraints
 \eq{constr-P}.

 The Hamiltonian of the system which is
equivalent one given by  \eq{S1} is 

\be
H = \l \hat \phi_1 + \l_1 \hat \phi_2,
\la{Ham}
\ee

where $\l$ and $\l_1$ are Lagrange multipliers.

The BRST charge is given by 

\be
Q =  c (P_1 + b X^{-}) + c_1 (P^{+} + b X^1)
\la{QP}
\ee

where $c$ and $c_1$ are ghosts corresponding to each first class constraint.
  It is also possible to include information that boundary action comes
 from string as was done in \cite{Wit1}, but the main result that conversion together
 with change of variables and fields redefinition gives ordinary, commutative behavior
 for the string boundary.
In some sense the conversion procedure is equivalent to restoring gauge symmetry of theory where
it was broken. For example if one starts from Maxwell theory with fixed gauge and then  introduce
new variables, that were fixed, i.e. nondynamical before, then use conversion, one get Abelian gauge
symmetry on the level of effective action.
And two models, with fixed gauge and with gauge invariance, are equivalent to each other not only on classical but
also on quantum level. In our case of interest the situation is almost the same except that we start from
theory without apparent gauge fix, but this theory could be interpreted as one with fixed gauge and the 
main task of conversion is just to find equivalent extended and therefore more general theory described
 by the system of first class constraints only.
 This procedure helps us to argue that noncommutativity appearing in initial theory is not a fundamental property
of the theory but just a some particular choice of gauge fixing of equivalent model.\\
 It is possible to conduct conversion for string in decoupling limit in more covariant way.
To do so let us start from system of constraints \eq{constrP} and then
 introduce new variables $\x^a$ and $P^\x_a$, $[\x^a, P^\x_b] = \d^a_b$ where $a = 1,2$.
Then generalizing \eq{P+} define

\be
P^+_a = P_a + P^\x_a, \qquad X^{+a} = X^a + \x^a,
\la{++}
\ee

\be
P^-_a = P_a - P^\x_a, \qquad X^{-a} = X^a - \x^a,
\la{--}
\ee

 Then \eq{constrP} could be modified into

\be
\hat \phi_a = P^-_a + b \e_{ab} X^{+b}.
\la{-+}
\ee 

Now all the constraints $\hat\phi_a$ are of the first class, i.e. commute. But, in difference from
previous example of conversion, degrees of freedom counting tells us that we need to add one more 
first class constraint, otherwise modified system is not going to be equivalent to initial one.
 This constraint could be chosen in the form

\be
\Psi = P^+ X^+ - P^- X^-.
\la{-+2}
\ee

This choice in some sense reminds constraint that could be obtained
from \eq{constrP} by projecting by $X^a$. If one puts all new variables equal to zero the constraints
\eq{-+} transform to \eq{constrP}, and \eq{-+2} is becoming equal to zero, i.e. does not dependent
on $\phi_a$ and doesn't carry any new information.

In this Section we gave two examples of conversion that produced equivalent systems 
with commuting variables (i.e. commuting string end-points). Those examples could be of the strong suggestion that
noncommutativity
could be removed not only on the level of constraints/Dirac brackets, but also on the level of two-point
functions.

\bigskip

\section { String in constant B-field, general case.}

Here we step aside from decoupling limit of string which was described in previous section and analyze
complete set of constraints coming from the action \eq{S1}.
It is more convenient to start not from Lagrangian \eq{S1} but rather from one without B-dependent
term \cite{Wit1} and impose boundary condition \eq{bound1} as additional boundary constraint.
The starting action is

\be
S = \frac {1}{4 \pi \a'} \int d^2 \s \del X^\m \del X_\m,
\la{S2S}
\ee

and boundary constraint

\be
X'^a + \dot X^b B_b{}^a = 0, \qquad X'^{m} = 0,
\la{X'S}
\ee

where $X^\m$ is full set of target-space coordinates for string, $X^a$ are
coordinates of string endpoints on the D-brane and $X^m$ is the rest of coordinates. 

As usually, we have first class constraints which follows from \eq{S2S} 

\be
H = 2 \pi \a' P^2 - \frac {1}{2 \pi \a'}  X'^2, \qquad H_1 = P_\m X'^\m.
\la{HamS}
\ee

The Hamiltonian is given by 

\be
H_t = \int d\s \biggl ( NH + N_1 H_1 \biggr).
\la{Hamiltonian}
\ee

The momentum in this case is 

\be
\dot X^a = 2 \pi \a' P^a.
\la{P2S}
\ee

Then the boundary condition \eq{X'S} could be rewritten as constraint

\be
\Phi^a  =  X'^a + 2 \pi \a' P^c B_c{}^a.
\la{Phi2S}
\ee

If B-field is absent, the boundary condition $X'=0$ leads to infinite number of constraints 
in the form 

\be
N^{(2k+1)} = 0, \qquad N_1^{(2k)} = 0, \qquad  X^{(2k+1)} = P^{(2k+1)} = 0,
\la{N}
\ee

where $(k)$ denotes k's derivative in respect to $\s$. All these constraints are
equivalent to extending $\s$ to $ [-\pi, \pi] $ and taking the orbifold projection
\cite{BH}, \cite{Lee} 

\be
X(- \s) = X(\s), \qquad P(-\s) = P(\s), \qquad N(-\s) = N(\s), \qquad N_1 (-\s) =- N_1(\s).
\la{orb}
\ee

In the presence of constant B-field the secondary constraints appear, as usually, from the fact that 
 commutator of $\Phi^a$  with Hamiltonian gives either constraint or condition for Lagrange
multipliers. First of all, in constant B-field, conditions for the Lagrange multipliers are
the same as in \eq{N} and using the linear combinations of $\Phi$'s leads to the same boundary
conditions for $X^a$ and $P_a$ as in \eq{N} except that for $X^a$ if $k=0$ we have constraint \eq{Phi2S}.

Because we start from \eq{S2S} but not \eq{S1} in difference from 
 \cite{Chu2}, \cite{Jab2}, \cite{Oh1} and \cite{Lee} we don't have modification of the
higher derivative constraints but rather after going to Dirac brackets contribution to the 
noncommutativity is given only due to $\Phi^a$  i.e.

\be
[X^a (\s), X^b(\s') ]_D = [X^a, X^b]  - \int d \s'' [X^a, \Phi^c (\s'')] C^{-1}_{cd} [P'^d(\s''),
X^b],
\la{D2S}
\ee

where $C^{-1}_{cd}$ is inverse matrix of commutator coefficients between $\Phi$ and $P'$, 
and also we have to use regularization for the endpoints (see \cite{Chu2}, \cite{Jab2}) , i.e.
for $\s, \s' = 0$ or $  \pi $. The same contribution is given by considering higher odd derivatives
of $P_a$.
We see that coordinates of endpoints do not commute but momenta do.\\
Let us investigate the nature of this noncommutativity. Here we will not use conversion of the system of
second class constraints into the first ones. It is more transparent to modify the system by the way that
second class constraints $\Phi, X^{(2k+3)} $ and $ P^{(2k+1)}$ reminded ones for the case of zero
B-field. The noncommutativity on the level of Dirac brackets appears because of the nonzero commutator
of $X^a$ and $\Phi^b$. If B-field is zero we see that they commute and Dirac brackets are the same as
Poisson
ones. Let us modify the system of constraints to obtain one that is equivalent to initial system. For a moment we
will
forget about higher derivative constraints, but it is straightforward to incorporate them into the
whole picture.
We will start from

\be
\Phi^a = X'^a + 2 \pi \a' P_c B^{ca}, \qquad P'^a = 0,
\la{Phi2SP}
\ee

where a runs from 1 to r. We have 2r second class constraints here.
To modify them introduce 2r new canonical variables $c^a$ and $P^{(c)}_a$, $[c^a, P^{(c)}_b] = \d^a{}_b$ 
and 2r additional second class
constraints. With modified initial ones we have

\be
\hat \Phi^a = X'^a + c^a, \qquad \phi_1^a = c^a - 2 \pi \a' P_c B^{ca},
\nonumber
\la{phi1S}
\ee

\be
P'_a = 0, \qquad P^{(c)}_a = 0.
\la{PPS}
\ee

Now, instead of 2r constraints we have 4r ones but we added 2r new variables, so
 number of degrees of freedom remains the same. The Dirac brackets \eq{D2S} are identically
equal to zero because $\hat \Phi$ commute with $X^a$ as in the case of string without B-field and
the rest of constraints \eq{phi1S}, \eq{PPS} also don't give any contribution to Dirac brackets.
Procedure, which we used here, is different from conversion. First of all because conversion assumes that
one can obtain initial constraints after fixing some particular values of additional variables.
For example in Section 2 we showed that after putting all new variables equal to zero one has initial
constraint system. Here it is not the case. We can't fix c, otherwise it gives $P=0$.
One can ask why new model is equivalent to old one. It could be argued that what we did is just
redefining the variables. And we don't have to fix any particular value
of $c^a$ or $P^{(c)}_a$ because additional constraints are not of the first class, but rather second one
and that's why they can be solved algebraically to produce initial system.\\
 Here we considered simplified system comparing to one of \cite{Chu2}. The main difference in sets of 
constraints
is that in the later case there were taken even higher derivatives of $\phi^a$ rather then $X'^a$.
And application of the same technique is straightforward.\\ 
  Now we see that introducing new variables even without changing the nature of constraints (from 
second to the first class) produces the
 Dirac brackets for string endpoints which are equal to zero.
 Therefore we were able to show that endpoints of the string expressed in new variables
 commute between each other. Analysis of two-point functions is not so straightforward as
 for the case of the decoupling limit but after change of variables and 
  field redefinition $< \hat X^a, \hat X^b >$,  where $\hat X$ are new variables
  playing role of effective string coordinates on the boundary, two-point functions become symmetric and
interpreting
time ordering as operator ordering gives commutativity.\\
  Here we argue that noncommutativity of string endpoints is not fundamental but rather
removable and depends on the change of basis. It also possible as in the Section 2 to consider modified 
BRST charge and perform analysis of mode decomposition  and to show that by the same way it is possible
to introduce new algebraic variables which produce commutativity of the string endpoints on the
level of Dirac brackets. It is becoming straightforward if one starts from constraints for the string
modes given in \cite{Lee} and modify them by introducing c and $P^{(c)}$ by the same way as in 
\eq{phi1S}.

\bigskip


\section{ Membrane in constant C-field, decoupling limit. }

In this section we discuss how to remove noncommutativity for the decoupling limit of 
membrane
ending on the M-5 branes. First of all, there are some crucial differences between taking 
decoupling 
limit of the string and membrane \cite{Berg1}, \cite{KS}. The only constant in eleven dimensions is the Plank
constant.
 Also we can't take flat background metric generated by the  five-branes as was was 
 explained in \cite{Berg1}. 
Moreover we have to consider stock of  five branes and probe membrane ending on one of the
five-branes. Then the decoupling limit could be found from the following :\\
  1.Bulk modes of the membrane must decouple and disappear.\\
2.String, that lives on the boundary, i.e. on the surface of the five-brane is fully described
by only Wess-Zumino membrane term.\\
3.Then, after decoupling, all dynamics of the world-volume theory of five-brane is governed
by $(2,0)$ six-dimensional tensor multiplet.\\
  As was shown in \cite{Berg1} and \cite{Berg2} by rewriting membrane kinetic term in terms of
Lorenz harmonics it is possible to produce decoupling explicitly in the limit when $l_p \ra 0$ but 
open membrane metric is fixed. Following \cite{Berg1} one can find convenient form of Wess-Zumino membrane term
in the case of constant C-field.
The Wess-Zumino term for membrane has two contributions - from pullback of eleven-dimensional three
form $A^{(3)}$ to membrane surface and from pullback of five-brane two-form $B^{(2)}$ to membrane
boundary,
i.e.

\be
S_{WZ} = \int_{M^3} A^{(3)} + \int_{\del M^3} B^{(2)}.
\la{CBM1}
\ee

The constant C-field on the five-brane is given by $C^{(3)} = dB^{(2)} + A^{(3)}$, where here
$A^{(3)}$ is pullback to five-brane world-volume. The notion of nonlinear self-duality for C-field
on the five-brane is extremely important here \cite{ES1}.
At least locally we can write $A^{(3)} = da^{(2)}$. Excluding part of $A^{(3)}$ which gives zero pullback
to the five-brane world-volume and using the fact that C is constant one has

\be
C^{(3)}_{\m\n\l} X^\l = 3(B^{(2)}_{\m\n} + a^{(2)}_{\m\n}).
\la{CBM}
\ee

The components of constant C-field could be chosen in the form \cite{Wit1}

\be
C^{(3)}_{012} = - \frac {h}{\sqrt{1+l_p^6 h^2}}, \qquad C^{(3)}_{456} = h.
\la{ChM}
\ee

The action of the membrane ending on M-5-brane in decoupling limit is given by two terms \cite{Berg1}

\be
S = \int d^2 \s \frac {h}{3 \sqrt{1+l_p^6 h^2}} \e_{ijk} X^i \dot X^j X'^k + \int d^2 \s \frac {h}{3}
\e_{abc} X^a \dot X^b X'^c,
\la{a1-brane}
\ee

where 5-brane world-volume was decomposed into the two three-dimensional  pieces, where
$X^i$ lies on the membrane and $X^a$ are  five-brane coordinates normal to the membrane 
and prime denotes differentiation in respect to $\s$ and dot - in respect to time.

Consider the second part of the \eq{a1-brane}

\be
S = \int d^2 \s \frac {h}{3} \e_{abc} X^a \dot X^b X'^c,
\la{a-brane}
\ee

The momentum is given by

\be
P_a = - \frac{h}{3} \e_{abc} X^b X'^c.
\la{mP}
\ee

With constraints

\be
\phi_a = P_a + \frac{h}{3} \e_{abc} X^b X'^c,
\la{mc}
\ee

and Poisson brackets

\be
[ \phi_a (\s), \phi_b (\s') ] = {2 \over 3}h \e_{abc} X'^c \d (\s - \s').
\la{mPhPh}
\ee

Among three constraints two are of the second class and one is the first class.
To see that project $\phi_a$ by $X, X', P$. Then first class constraint is

\be
\Psi = X'^a P_a,
\la{mc1}
\ee

and the second ones

\be
X^a P_a = 0, \qquad P^2 + {h \over 3} \e_{abc} P^a X^b X'^c = 0.
\la{mc2}
\ee

The equations of motion are 

\be
\e_{abc} \dot X^b X'^c = 0.
\la{EMM}
\ee

Here, because we have a mixture of first and second class constraints it is more convenient to 
convert them directly in the mixture, i.e. we need to find additional variables which give us all
constraints of the first class. The discussions on conversion in mixture of first and second
class constraints and references therein could be found in \cite{BMRS}.
 To do it covariantly in three dimensions orthogonal to membrane
world-volume let us introduce new variables $\x^a$ and $P^{(\x)}_b$ with 
$[\x^a (\s),P^{(\x)}_b (\s')] = \d^a_b \d (\s - \s')$. But if we want to convert two second
class constraints into the first one counting of degrees of freedom tells us that we need only two extra
variables and
we added six. The resolution of this problem is that not all of the above new variables are
independent and that's why we need to impose two extra first class constraints that kills four degrees
of freedom leaving as with only two extra independent variables that we needed.
 It is convenient to proceed as follows. Let us define
 
\be
  X^{a+} = X^a + \x^a, \qquad P^{a+} = P^a + P^{(\x) a} , 
\la{XP+}
\ee
 
\be
  X^{a-} = X^a - \x^a, \qquad P^{a-} = P^a - P^{(\x) a} . 
\la{XP+1}
\ee

Then the modified constraints \eq{mc} could be rewritten as the following first class constraints

\be
\hat \phi_a = P_{a}^{-} + \frac {h}{3} \e_{abc}X^{b+}X'^{c+},
\la{HPhiM}
\ee

and they identically commute.
Now we need two additional commuting with everything constraints.
It is possible to choose them in the form

\be
P^+ X^+ - P^- X^- = 0 ,  \qquad  P'^+ X^+  + P^- X'^-= 0.
\la{NewCM}
\ee

It is not unique but convenient choice. Now all the constraints \eq{HPhiM} and \eq{NewCM}
are of the first class.
The BRST charge is now given by sum of the first class constraints multiplied by corresponding ghosts
plus contribution from commutators.

This BRST charge describes theory equivalent to initial one not only on the classical but 
also on the quantum level. Fixing gauge $\x=0$ and $P_\x = 0$ gives initial theory.
In some sense one can think that model before conversion was one with fixed gauge symmetries
, which could be restored on the level of BRST invariant action. By the same way one can proceed
for the first term in \eq{a1-brane}.

In this example we see that even for the nonlinear case of membrane boundary it is possible
to change variables to end up with commutative coordinates. This procedure is close to one in the
end of second Section except that for the case of membrane one of the constraints was of the
first class and we were forced to have conversion directly in the mixture of two types of
constraints.

\bigskip
\section{Discussions and Conclusion.}

In this note we showed that by appropriate change of variables and extension
 of the phase space of the models of
noncommutative string and membrane they appear to be equivalent to ordinary ones. 
It is possible to argue that noncommutativity of string/membrane endpoints,
 appearing fundamental, 
is rather removable and depends on field content and change of variables.
Not only conversion is doing it's job, as for the case of string/membrane in 
decoupling limit, but also
equivalence  of two phase spaces without changing type of constraints is giving the 
same result. It is important to investigate the correspondence between noncommutative
string and ordinary one from worldsheet point of view. It is not quite clear yet
how this equivalence works for the product of vertex operators, which apriori obey
noncommutative algebra.

 Here we presented only analysis of bosonic part of theories. 
It is interesting to consider whole supersymmetric
case extending  approach of \cite{CZ}, \cite{LW}, \cite{LZ}.
For supersymmetric case it is also important to understand situation on the level
of supersymmetric solutions and Dp-Dq brane systems \cite{MPT},\cite{Wit4}.

 Hamiltonian systems with boundaries were also extensively studied in \cite{Z} 
 from the similar point of
view as well as in \cite{B1}, \cite{Sol} where instead of interpreting 
boundary conditions
as constraints the Poisson brackets were modified to include boundary contribution. 
It is interesting to compare results coming from two approaches.
 
 It could be noted that membrane boundary string in decoupling limit and 
 critical C-field should be not only
 tensionless but also doesn't have gravity in it's spectrum, that reminds a Little String 
 Theory concept, see \cite{TH} and references therein. 
 Usual world-sheet formulation of tensionless string \cite{T1}  has kinetic term quadratic
in fields together with null vectors and world-sheet of such a string is a null surface.
Sometimes in literature it is called null-string.
 Unfortunately  this theory is anomalous and complete spectrum is not known yet.
It could be interesting to assume
 that tensionless string could be described by equations of motion \eq{mc}
and by Lagrangian \eq{a-brane}. But because \eq{a-brane} describes membrane boundary
in decoupling limit the mentioned above tensionless string doesn't contain graviton
in it's spectrum and it's low energy limit is described by $(2,0)$ six-dimensional theory.
Therefore there is a connection to Little String Theory.
Modified system of  the first class constraints \eq{HPhiM} and \eq{NewCM}
could be quantized in the BRST framework. At least it is straightforward to
conduct operator BRST quantization. It is interesting to find a spectrum 
of this model as well as build vertex operators and study this model from world-sheet 
point of view.

It is interesting to ask what are the other nonlinear models which posses the same 
equations of motion
 as \eq{EMM} and carry some resemblance with the membrane in the decoupling limit.
 
  First of all, because action \eq{a-brane} is pure Wess-Zumino term of the
membrane it is useful to start from known nonlinear Wess-Zumino models in two dimensions.
Let us consider Wess-Zumino term of $SU(2)$ WZNW model. It is given by the action:

\be
S_{WZ-WZNW} = \frac{k}{96 \pi} \int dr \int  d^2 \x g^{-1} \dot g  g^{-1} \e^{\m\n}\del_\m g
g^{-1} \del_\n g.
\la{WZWG}
\ee

Using  parametrization of $SU(2)$ in terms of Euler`s angles $\phi, \theta, \psi$ it
 is possible to rewrite \eq{WZWG} as

\be
S_{WZ-WZNW} = \frac {k}{4 \pi} \int d^2 \x \phi \e^{\m\n}  sin\theta \del_\m \theta 
\del_\n \psi.
\la{WZSU2}
\ee

At least for compact $ X^a$ it is possible to identify $(\phi, -cos\theta, \psi)$ with
 $(X^3, X^4, X^5)$
consequently. In this case WZ part of WZNW $SU(2)$ model could be rewritten as

\be
S_{WZ-WZNW} = \frac {k}{4 \pi} \int d^2 \x X^3(\dot X^4 X'^5 - \dot X^5 X'^4),
\la{WZWX}
\ee

and it is always possible to go to free-field realization of \eq{WZWX}.
This action gives the same equations of motion as \eq{EMM}. So, two models, string 
describing membrane endpoints on five-brane  and
Wess-Zumino part of $SU(2)$ WZNW model, are equivalent on the level of equations of motion,
 but not the action because the symmetries of Lagrangians are different. Therefore,
it could be useful to find a connection if any  between description of   
evolution of the boundary string on the surface of five-brane and 
 topological part of $SU(2)$ WZNW model at least for compact 
coordinates of membrane end-points.



\bigskip \bigskip \noindent{\large \bf Acknowledgments.}

\medskip

 I would like to thank  A.Sadrzadeh, T.A.Tran for discussions and especially
thank E.Sezgin for fruitful discussions and for constant encouragement.

\bigskip


\newpage


\end{document}